\documentclass[twocolumn,showpacs,superscriptaddress, preprintnumbers,amsmath,amssymb,prl]{revtex4}
\usepackage{latexsym}
\usepackage{epsfig}
\usepackage{graphics}
\usepackage{color}
\usepackage{times}
\usepackage{pst-all}
\usepackage{draftcopy}
\draftcopyFirstPage{1} 
\draftcopyLastPage{2}
\draftcopySetGrey{1}

\begin{document}
\title{Spurious detection of phase synchronization in coupled
nonlinear oscillators}
 \author{Limei Xu$^{1}$, Zhi Chen$^{1,2}$, Kun
Hu$^{1,3}$, H. Eugene Stanley$^{1}$, and Plamen Ch. Ivanov$^{}$}

\affiliation{$^{}$Center for Polymer Studies and Department of Physics, Boston
  University, Boston, MA 02215 USA \\
$^{2}$Department of Physics and Astronomy, University of California, Irvine,
California 92697, USA\\
$^{3}$Division of Gerontology, Harvard Medical School, Beth Israel Deaconess
Medical Center, Boston, Massachusetts 02215, USA }

\date{\today}


\begin{abstract}
Coupled nonlinear systems under certain conditions exhibit phase
synchronization, which may change for different frequency bands or with
presence of additive system noise. In both cases, Fourier filtering is
traditionally used to preprocess data. We investigate to what extent the
phase synchronization of two coupled R\"{o}ssler oscillators depends on (1)
the broadness of their power spectrum, (2) the width of the band-pass filter,
and (3) the level of added noise. We find that for identical coupling
strengths, oscillators with broader power spectra exhibit weaker
synchronization. Further, we find that within a broad band width range,
band-pass filtering reduces the effect of noise but can lead to a spurious
increase in the degree of synchronization with narrowing band width, even
when the coupling between the two oscillators remains the same.
\end{abstract}
\maketitle

\begin{figure*}
\includegraphics[width=10cm,height=10cm,angle=0]{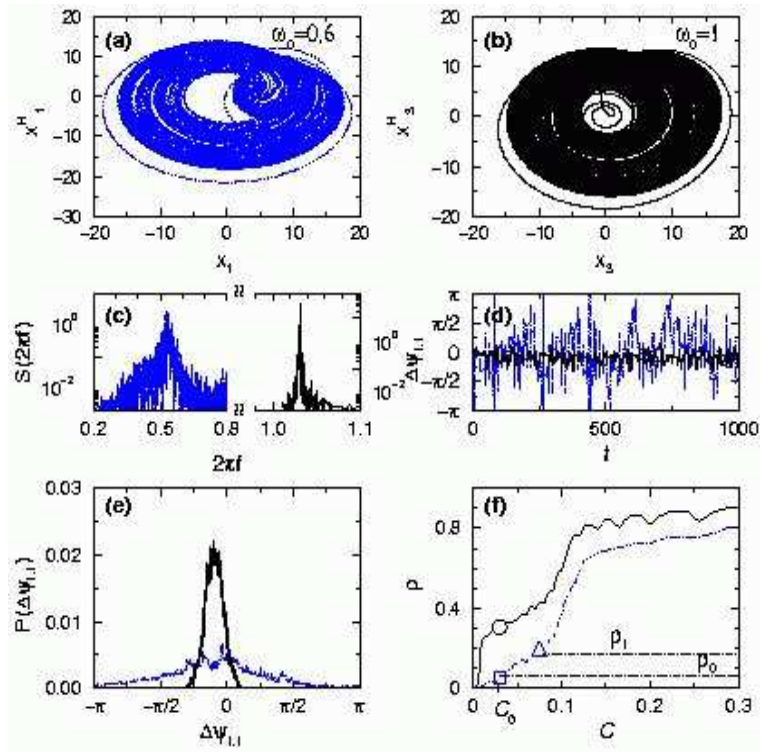}
\caption{Differences in the synchronization of two R\"{o}ssler systems with
    identical coupling strengths and different power spectra. Phase plot
    trajectories of the variables $x$ vs. their Hilbert transform $x^{H}$
    for: (a) system (1,2), with $x_{1}$ corresponding to
    $\omega_{1}=\omega_{0}+\Delta\omega$, where $\omega_{0}=0.6$ and
    $\Delta\omega=0.005$; (b) system (3,4), with $x_{3}$ corresponding to
    $\omega_{3}=\omega_{0}+\Delta\omega$, where now $\omega_{0}=1$ and
    $\Delta\omega=0.015$. For both R\"{o}ssler systems $C=0.03$. (c) Power
    spectra of the time sequence $x_{1}$ (dashed line) and $x_{3}$ (solid
    line). A broader spectrum is observed for system (1,2) compared to system
    (3,4). (d) Instantaneous phase difference $\Delta\psi_{1,1}\equiv
    (\phi_{x_{1}(t)}-\phi_{x_{2}(t)}) \mod (2\pi)$ for system (1,2) (dashed
    line), and $\Delta\psi_{1,1}\equiv (\phi_{x_{3}(t)}-\phi_{x_{4}(t)}) \mod
    (2\pi)$ for system (3,4) (solid line), and (e) their corresponding
    distributions $P(\Delta\psi_{1,1})$. System (1,2) exhibits larger
    fluctuations in $\Delta\psi_{1,1}$ and is characterized by a broader
    distribution $P(\Delta\psi_{1,1})$. (f) Synchronization index $\rho$ as a
    function of the coupling strength $C$. For identical values of $C$,
    system (3,4) (solid line) which is characterized by a narrower power
    spectrum exhibits stronger synchronization (larger index $\rho$) compared
    to system (1,2) with a broader power spectrum. Specifically, for
    identical coupling strength $C=C_{0}=0.03$, the index $\rho=\rho_{0}$
    ($\Box$) for system (1,2), while $\rho=0.3>\rho_{0}$ ($\circ$) for system
    (3,4) although the frequency mismatch for system (3,4) is much
    larger. The effect of a Fourier band-pass filter applied to the system
    (1,2) while keeping $C=0.03$ fixed is equivalent to an increase of the
    coupling strength of the system leading to a larger index
    $\rho_{1}>\rho_{0}$ ($\bigtriangleup$) as also shown in
    Fig.~\ref{fig2}(e).}
\label{fig1}
\end{figure*}

\begin{figure}
  \includegraphics[width=7cm,angle=0]{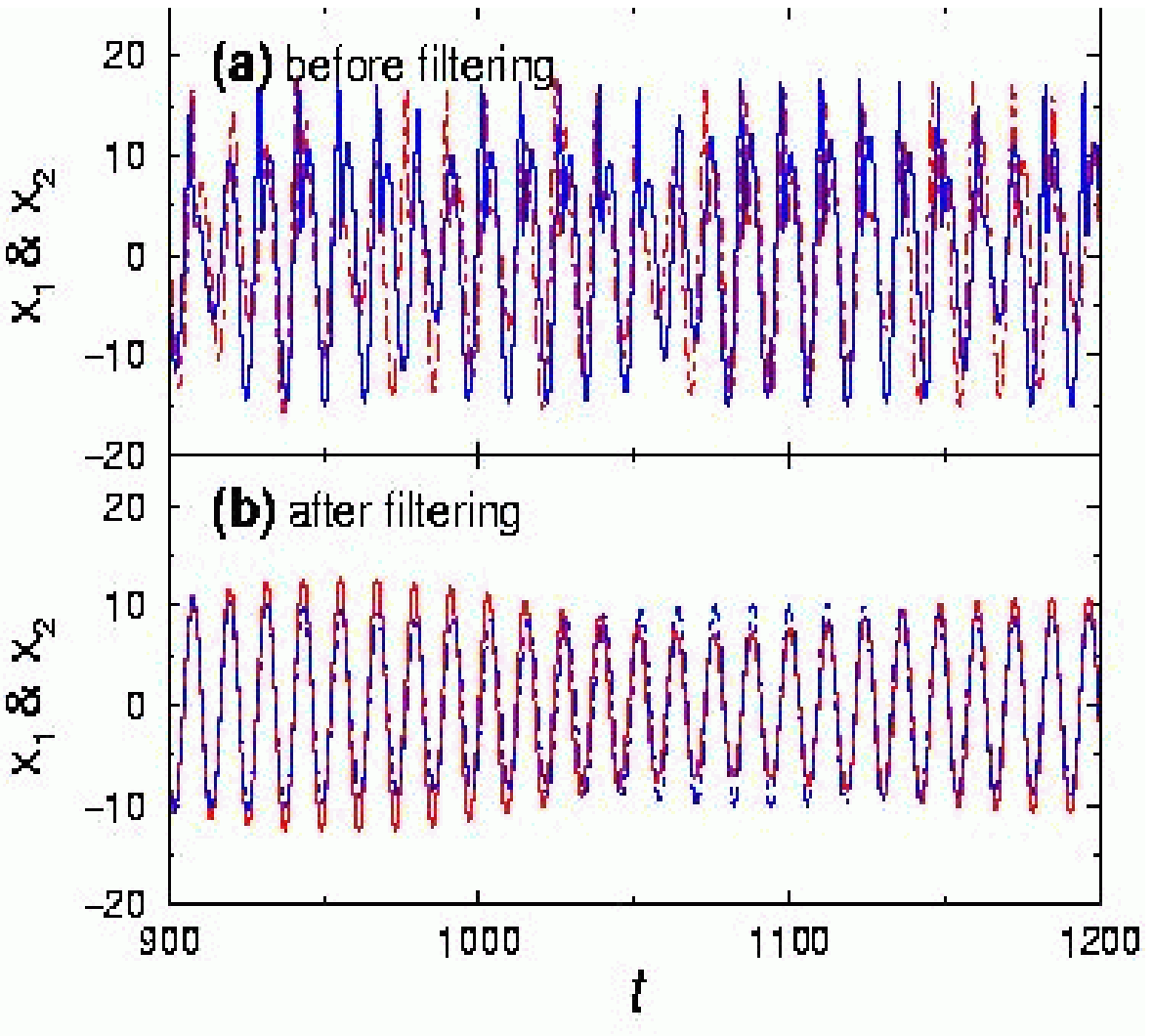}
  \includegraphics[width=7cm,angle=0]{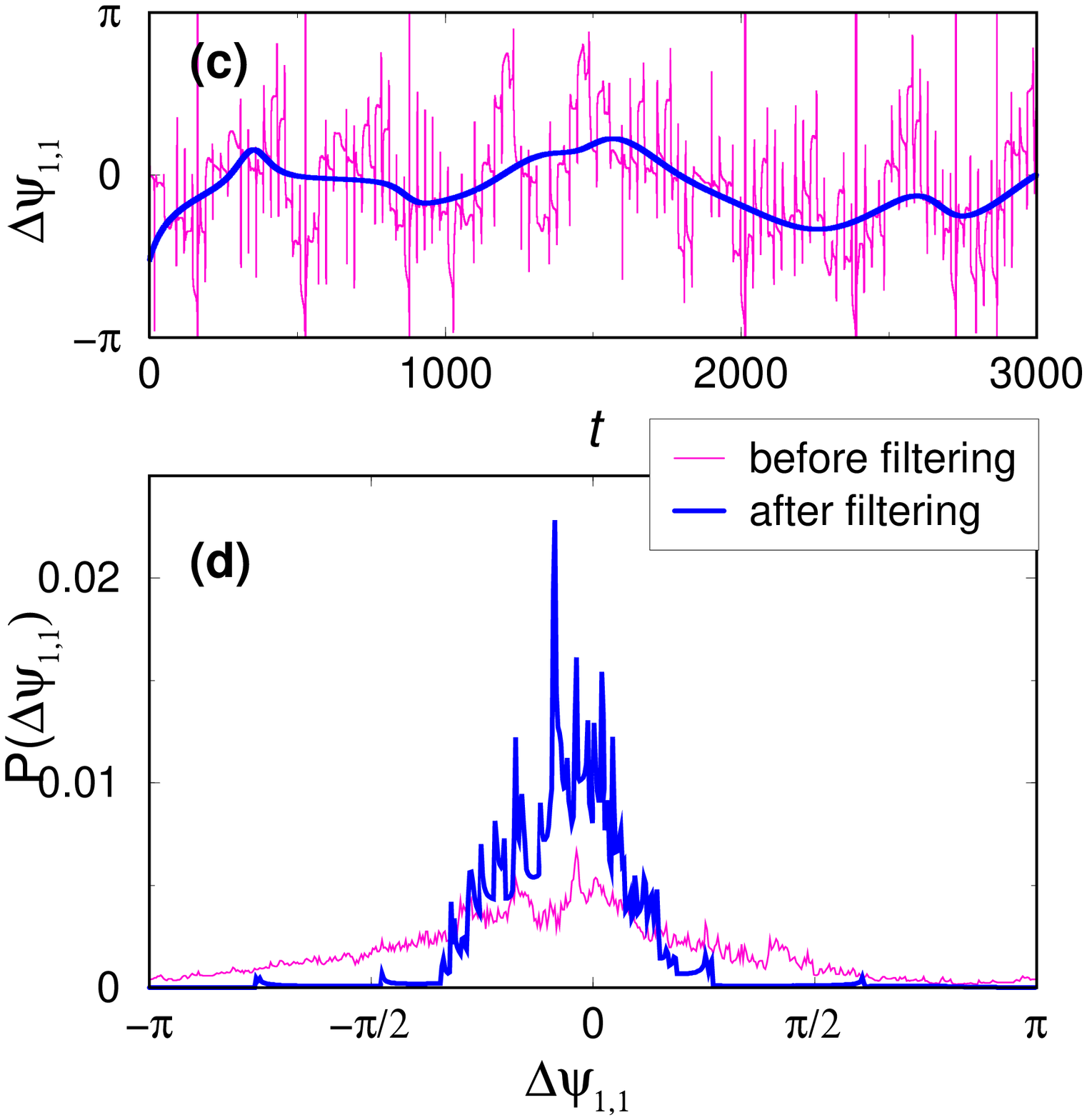}
  \includegraphics[width=7cm,angle=0]{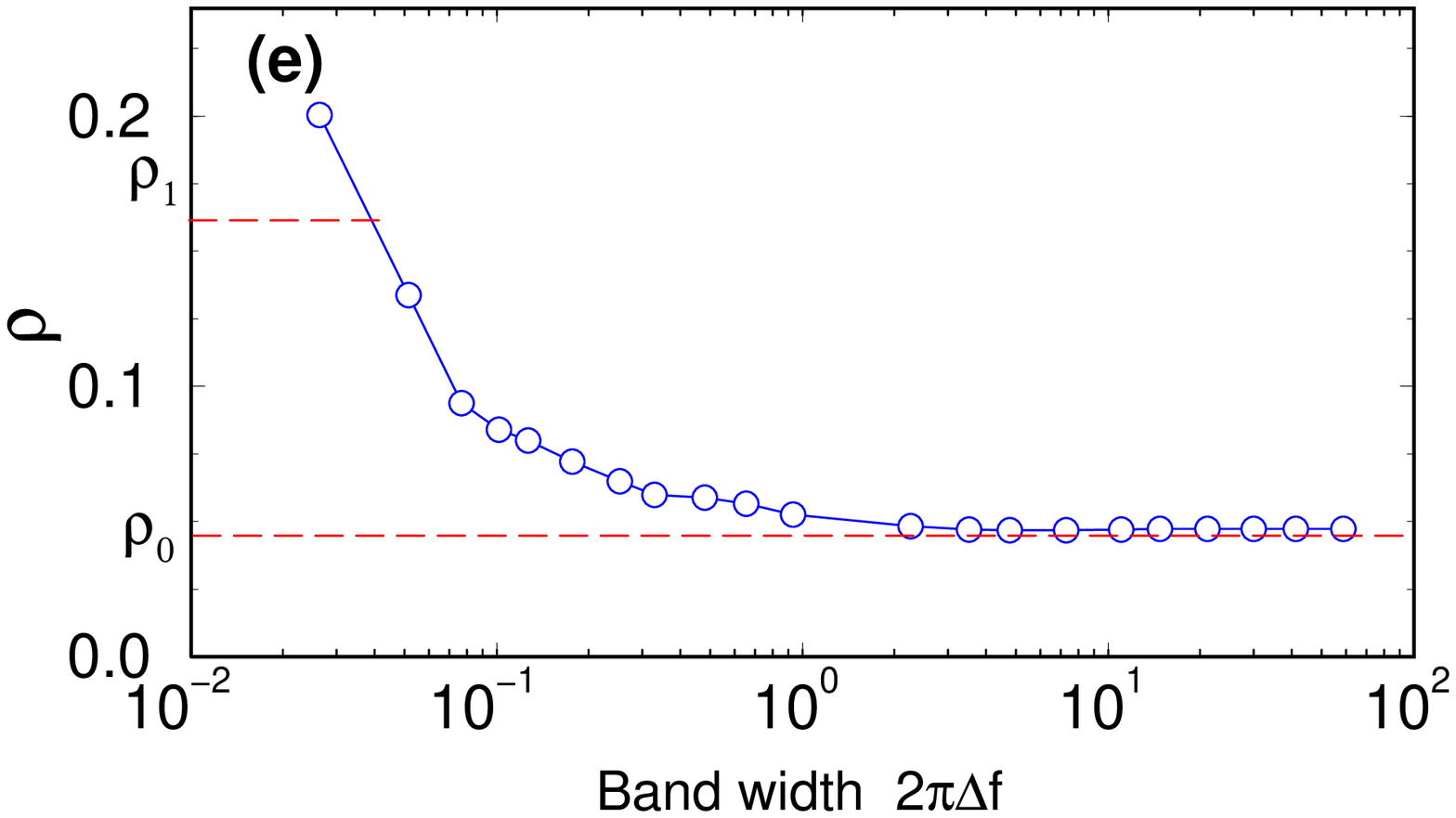}
  \caption{Effects of band-pass filtering on synchronization. Time sequence
of the variables $x_{1}$ and $x_{2}$ of system (1,2): (a) before and (b)
after applying a band-pass Fourier filter with band width $\Delta
f=0.01$.
 After band-pass filtering the
sequences $x_{1}$ and $x_{2}$ are better aligned in time (with almost
matching peaks). (c) Instantaneous phase difference $\Delta\psi_{1,1}$, and
(d) the distribution $P(\Delta\psi_{1,1})$ before (dashed line) and after
(solid line) the Fourier band-pass filtering. After filtering,
$\Delta\psi_{1,1}$ is characterized by less fluctuations and a much narrower
distribution $P(\Delta\psi_{1,1})$, indicating a stronger synchronization,
although the coupling strength $C=0.03$ remains constant. (e) Dependence of
the index $\rho$ on the band width $2\pi\Delta f$ for fixed
$C=0.03$. A filter with a relatively broader band width ($2\pi\Delta f>1$)
leaves the synchronization index $\rho$ practically unchanged,
$\rho=\rho_{0}$, where $\rho_{0}$ characterizes the synchronization between
$x_{1}$ and $x_{2}$ before filtering. Narrowing $\Delta f$
leads to a sharp increase in $\rho$, which is an artifact of the
Fourier filtering as the coupling $C$ and all other parameters remain
unchanged, e.g, for $\Delta f=0.005$, $\rho=\rho_{1}\approx 4\rho_{0}$.}
  \label{fig2}
\end{figure}

\begin{figure}
\includegraphics[width=7cm, height=8cm]{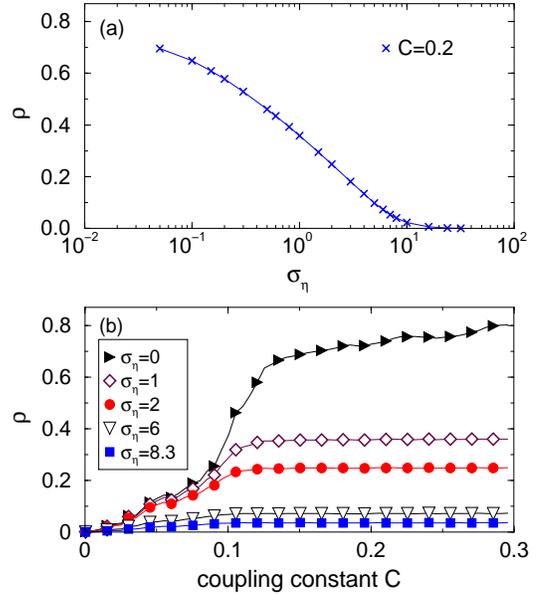}
\caption{Effect of external additive white noise on phase synchronization for
system (1,2). (a) Dependence of the synchronization index $\rho$ on the noise
strength $\sigma_{\eta}$ for fixed value of the coupling constant $C$. (b)
Dependence of the synchronization index $\rho$ on the coupling strength $C$
for different levels of white noise which are defined through the standard
deviation $\sigma_{\eta}$. }
\label{fig3}
\end{figure}

\begin{figure}
\includegraphics[width=8cm,angle=0]{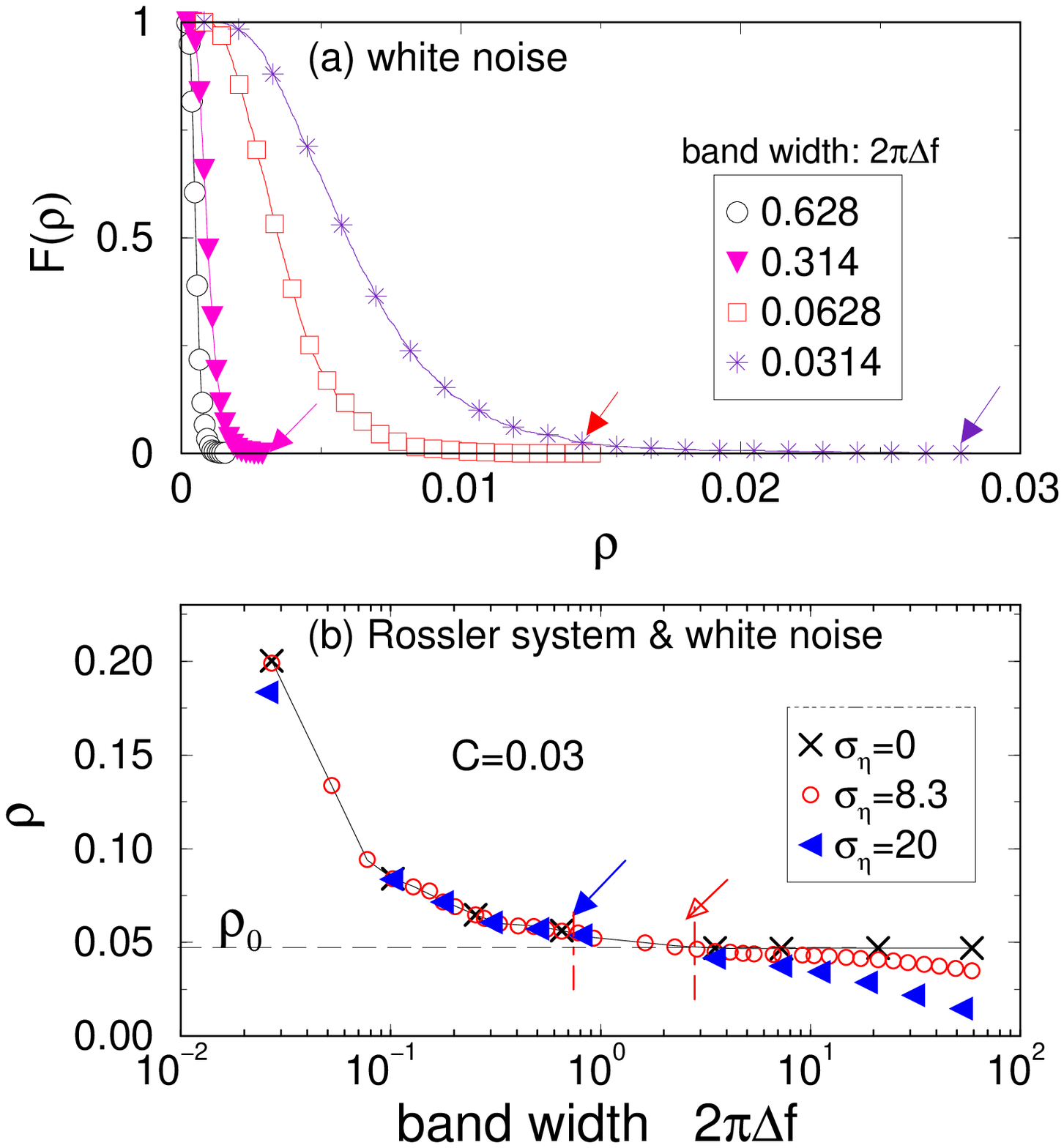}
\caption{Combined effects of external noise and Fourier band-pass filtering
on the synchronization. (a) Cumulative distribution function $F(\rho)\equiv
1-\int_{0}^{\rho} P(\rho{'}) d\rho{'}$ for the index $\rho$ obtained from
$100$ different realizations of pairs of white noise signals without
coupling. The length of the noise signals is $int[10^7/2\pi]$. Tails of the
distributions for each band width indicate the maximum values of $\rho$ one
can obtain simply as a result of band-pass filtering when there is no
synchronization between two white noise signals. (b) Synchronization index $\rho$
obtained for system (1,2) with additive white noise as a function of the band
width $\Delta f$ for $C=0.03$. While the effect of noise is gradually reduced
by the Fourier band-pass filter with decreasing band width $\Delta f$, there
is an artificially increased synchronization (sharp increase in $\rho$) when
$2\pi \Delta f<1$, as also shown in Fig.~\ref{fig2}(e).}
\label{fig4}
\end{figure}

 In recent years both theoretical and experimental studies of coupled
nonlinear oscillators has demonstrated that such oscillators can exhibit
phase
synchronization~\cite{Rosenblum_PRL_1996,Rosenblum_PRL_1997,Parlitz_PRE_1996,Pikovsky_book_2001,Pikovsky_physicaD_1997}. Analysis
of experimental data has also indicated the presence of phase synchronization
in a range of coupled physical, biological and physiological systems
~\cite{Schafer_nature_1998,Anishcherko_int_J_Bif_chaos_2000,Stefanovska_PRL_2000,Bahar_chaos_2003,Rybski_physica_2003,Moshel_ANY_Acad_Sci_2005,Boccaletti_PRE_2000,13,Nagai_PRE_2005,Pereda_Neurobiology_2005,Boccaletti_phys_report_2002,
Zchen_PRE_2006}. In many of these studies, an important practical question is
how multi-variate time series characterized by relatively broad power
spectrum are phase synchronized in a specific frequency
range~\cite{Pierro_PRL_1998,Gross_J_physiology_2000,Quiroga_PRE_2002,Angelini_PRL_2004,Mormann_PhysicaD_2000,Neiman_PRL_1999,Gysels_Signal_processing_2005}. The
presence of internal or external noise may also be an obstacle when
quantifying phase synchronization from experimental
data~\cite{Pierro_PRL_1998,Neiman_PRL_1999,Zheng_2005,Lindner_Physics_report_2004,Gora_physica_2005}. In
both cases a band-pass filter is traditionally applied either to reduce the
noise effect or to extract the frequency range of interest. Thus, it is
important to know to what extent the width of the band-pass filter influences
the results of the phase synchronization analysis, as well as what is the
range of the index values obtained from the analysis that indicate a
statistically significant phase synchronization.

To address these questions, we consider a system of two coupled R\"{o}ssler
oscillators (1,2) defined as
\begin{eqnarray}
\label{EQ1}
\dot{x}_{1,2}&=&-\omega_{1,2} y_{1,2} - z_{1,2} + C (x_{2,1}-x_{1,2}),\nonumber
\\ \dot{y}_{1,2}&=&\omega_{1,2} x_{1,2} + a y_{1,2}, \nonumber \\
\dot{z}_{1,2}&=&f + z_{1,2} (x_{1,2}-b)
\end{eqnarray}
with parameter values $a=0.165$, $f=0.2$, and $b=10$. For the mismatch of
natural frequencies, we choose $\omega_{1,2}=\omega_{0}\pm\Delta\omega$, with
$\omega_{0}=0.6$ and $\Delta \omega=0.005$ [Fig.~\ref{fig1}(a)]. The time
step in our simulation is $\Delta t=2\pi/10^{3}$, and the signal length
$n=int[t/\Delta t]$ with $t=10^{4}$, where $int[x]$ denotes the integer part
of $x$.

  We first investigate the characteristics of the system defined in
Eq.~(\ref{EQ1}) by comparing them with the characteristics of a second set of
two coupled R\"{o}ssler oscillators (3,4) studied in
~\cite{Rosenblum_PRL_1997}.The system (3,4) is also described by
Eq.(\ref{EQ1}), and has the same values for parameters $a$, $f$, and $b$ as
system (1,2). The only differences are the natural frequency $\omega_{0}=1$
and the frequency mismatch $\Delta\omega=0.015$~[Fig.~\ref{fig1}(b)]. We
observe a significantly broader power spectrum for system (1,2)
with~$\omega_{0}=0.6$ and frequency mismatch
$\Delta\omega=0.005$~[Fig.~\ref{fig1}(c)]. Further, we observe that the
instantaneous phase differences
$\Delta\psi_{1,1}=[\phi_{x_{1}}(t)-\phi_{x_{2}}(t)] \mod (2\pi)$ for system
(1,2) exhibits larger fluctuations [Fig.\ref{fig1}(d)], described by a
broader distribution [Fig.\ref{fig1}(e)], compared to system (3,4),
suggesting a weaker 1:1 phase synchronization for system (1,2). To quantify
the degree of phase synchronization in the two R\"{o}ssler systems we use the
synchronization index $\rho=(S_{\rm max}-S)/S_{\rm max}$
\cite{Pierro_PRL_1998}, where $S\equiv-\sum_{k=1}^{N}P_{k}\ln P_{k}$ is the
Shannon entropy \cite{shannon_1949} of the distribution $P(\Delta\psi_{1,1})$
of $\Delta \psi_{1,1}$, and $S_{\rm max}=\ln N$, where
$N=int[\exp(0.626+0.4\ln(n-1.0))]$ is the optimized number of bins over which
the distribution is obtained~\cite{Otnes_1972}. For system (3,4) with a
narrow power spectrum we obtain a significantly larger value of $\rho$
compared to the system (1,2) characterized by a broader power spectrum
[Fig.\ref{fig1}(f)]. Varying the values of the coupling strength $C$, we find
that the phase synchronization index $\rho$ is consistently higher for system
(3,4) characterized by the narrower power spectrum. Thus, for the same
coupling strength $C$ and for identical other parameters, system (1,2) with
$\omega_{0}=0.6$, which has a broader power spectrum, exhibits weaker
synchronization compared to system (3,4) with $\omega_{0}=1$, which has a
narrow power spectrum. These findings are complementary to a recent study
indicating a different degree of phase synchronization for the spectral
components of coupled chaotic oscillators~\cite{Hramov_PRE_2005}.

Recent work has shown that coupled R\"{o}ssler oscillators may exhibit
different degrees of synchronization for different ranges of time scales
obtained via wavelet transform~\cite{Hramov_physciaD_2005}. Here, we ask to
what extent the width of a band-pass filter affects the degree of phase
synchronization between two coupled R\"{o}ssler oscillators. While the
output observables $x_{1}$and $x_{2}$ of system (1,2) are clearly not in
phase [Fig.\ref{fig2}(a)], after Fourier band-pass filtering in the range of
$\Delta f=0.01$ centered at the peak of the power spectrum $2\pi f\approx
0.54$ [Fig.~\ref{fig1}(c)], the observables $x_{1}$ and $x_{2}$ appear 1:1
synchronized with well aligned peaks [Fig.~\ref{fig2}(b)]. The effect of the
band-pass 
filter can be clearly seen in the behavior of the instantaneous
phase difference $\Delta \psi_{1:1 }$ [Fig.~\ref{fig2}(c)] and in the shape
of the probability density function $P(\Delta\psi_{1,1}(t))$
[Fig.\ref{fig2}(d)]. After band-pass filtering, $\Delta \psi_{1,1}$ becomes
smoother with less fluctuations, and the distribution $P(\Delta\psi_{1,1})$
exhibits a well pronounced peak. To quantify how the degree of
synchronization changes with the width $\Delta f$ of the band-pass filter, we
calculate the synchronization index $\rho$ [Fig.\ref{fig2}(e)]. We find that
for very large values of the band width $\Delta f$, the index $\rho$ is the
same as the value $\rho_{0}$ obtained for the system (1,2) without any
filtering, and that $\rho$ remains unchanged for intermediate values of
$\Delta f$. However, for decreasing $\Delta f$, the index $\rho$ increases
rapidly from the expected value $\rho_{0}$ [Fig.~\ref{fig1}(f), and
~\ref{fig2}(c)]. Such deviation to higher values of $\rho>\rho_{0}$, while
the coupling constant $C$ in Eq.~(\ref{EQ1}) remains fixed, indicates a
spurious effect of synchronization due to the band-pass filter. Thus,
applying a band-pass filter with a too narrow band width when pre-processing
empirical data may lead to overestimation of the phase synchronization (as
defined by index $\rho$) between two empirical systems where the coupling
strength is not known {\it a-priori}.

Many physical and biological systems are influenced by external noise, which
can mask their intrinsic properties. Recent studies have shown that noise can
bias the estimation of driver-response relationship in coupled nonlinear
oscillators leading to change in synchronization
measures~\cite{Grassberger_PRE_2000}. Specifically, external noise may weaken
the detection of the coupling and reduce the synchronization between two
coupled dynamical systems. To address this problem, we next test the effect
of external noise on the degree of phase synchronization of the two coupled
R\"{o}ssler oscillators defined in Eq.(\ref{EQ1}). Adding uncorrelated and
unfiltered Gaussian noise $\eta$ to the output observables $x_{1}$ and
$x_{2}$, while keeping the coupling constant $C$ in Eq.~(\ref{EQ1}) fixed, we
find that the synchronization index $\rho$ decreases with increasing noise
strength $\sigma_{\eta}$, (i.e., higher standard deviation $\sigma_{\eta}$
compared to the standard deviation $\sigma$ of the output signals $x_{1}$ and
$x_{2}$) [Fig.\ref{fig3}(a)]. The dependence of $\rho$ on the value of the
coupling constant $C$ for different noise strength is shown in
Fig.~\ref{fig3}(b). We find that the transition to the state of maximum
degree of synchronization (indicated by a horizontal plateau for $\rho$ in
Fig.~\ref{fig3}(b)) occurs at decreasing values of the coupling constant $C$
for increasing noise strength $\sigma_{\eta}$. For very strong noise
($\sigma_{\eta}=\sigma=8.3$), the two R\"{o}ssler oscillators in
Eq.(\ref{EQ1}) appear not to be synchronized, characterized by low values for
the index $\rho$, even for very large values of the coupling constant $C$
[Fig.~\ref{fig3}(b)]. We note, that with increasing noise strength
$\sigma_{\eta}$ the position of the crossover to the plateau of maximum
synchronization shifts to smaller values of $C$ in Fig.\ref{fig3}(b),
indicating that with increasing $\sigma_{\eta}$ the level of the plateau
drops faster compared to the decline in the growth of $\rho$ with increasing
coupling $C$.

To reduce the effect of noise in data analysis, a common approach is to apply
a band-pass filter. In the case of the coupled R\"{o}ssler oscillators
defined in Eq.(\ref{EQ1}), we ask to what extent a band-pass filter can
reduce the effect of external noise while preserving the expected ``true''
phase synchronization as presented by $\rho_{0}$ in Fig.~\ref{fig1}(e). To
answer this question, we first need to determine what are the limits to which
spurious phase synchronization can be obtained purely as a result of
band-pass filtering of two uncorrelated and not coupled Gaussian noise
signals. Our results for the synchronization index $\rho$ obtained from
multiple realizations of pairs of uncoupled white noise signals show that the
synchronization index $\rho$ can reach different maximum values $\rho_{max}$,
indicated by arrows in Fig.~\ref{fig4}(a), for different band width $\Delta
f$ --- with decreasing the band width $\rho_{\rm max}$ increases. The values
of $\rho_{\rm max}$ provide an estimate of the maximum possible effect
additive noise may have on the spurious ``detection~'' of phase
synchronization in coupled oscillators. Thus, empirical observations of
synchronization index $\rho>\rho_{\rm max}$ may indicate presence of a
genuine phase synchronization between the outputs of two coupled oscillators,
which is not an artifact of external noise. Our simulations show that the
value of $\rho_{max}$ does not change significantly with the length of the
uncorrelated noise signals. In Fig.~\ref{fig4}(b) we show how the
synchronization index $\rho$ for system(1,2) depends on the strength of the
added noise and on the width $\Delta f$ of the band-pass filter. For very
broad band width $\Delta f$ the noise is not sufficiently filtered, and the
synchronization between the two oscillators decreases ($\rho$ decreases) with
increasing noise strength $\sigma_{\eta}$. With decreasing band width $\Delta
f$, i.e., applying a stronger filter, the effect of the noise is reduced, and
correspondingly the index $\rho$ increases --- approaching the value
$\rho_{0}$ expected for the system (1,2) without noise. On the other hand,
applying a filter with too narrow band width $\Delta f$ leads to a spurious
synchronization effects with $\rho>\rho_{0}$ [Fig.\ref{fig4}(b)], following
closely the dependence of $\rho$ on $\Delta f$ shown in Fig.~\ref{fig2}(e)
for a R\"{o}ssler system without noise.

In summary, our results indicate that phase synchronization between coupled
nonlinear oscillators may strongly depend on the width of the power spectrum
of these oscillators. Further, we find that while external noise can affect
the degree of phase synchronization, band-pass filtering can reduce noise
effects but can also lead to a spurious overestimation of the actual degree
of phase synchronization in the system. This is of importance when analyzing
empirical data in specific narrow frequency ranges, for which the coupling
strength may not be known {\it a-priori}.

We thank NIH (Grant No. 2R01 HL 071972) for support.

\end{document}